\documentclass[12pt]{iopart}
\usepackage{graphicx}
\usepackage{bm}

\newcommand{\vek}    [1] {\textrm{\textbf{{#1}}}} 
\newcommand{\text}  [1]  {\mathrm{#1}}
\newcommand{\mi}         {\textrm{i}}   

\newcommand{\dd}     [0] {\text{d}}
\newcommand{\Op}     [1] {\hat{\textrm{#1}}}
\newcommand{\ket}    [1] {| #1 \rangle}
\newcommand{\bra}    [1] {\langle #1 |}
\newcommand{\bk}     [2] {\langle #1 | #2 \rangle}

\newcommand{\bok}    [3] {\langle #1 | #2 | #3 \rangle}
\newcommand{\Nst}    [1] {${}^1$#1}

\newcommand{\Nuk}    [4] {\mbox{${}_{#1}^{#2}\text{#3}^{\,#4}$}}

\newcommand{\He}         {\text{He}}

\newcommand{\HeH}        {\Nuk{}{}{HeH}{+}}
\newcommand{\au}{\text{a.u.}}
\newcommand{\eV}{\text{eV}}
\newcommand{\keV}{\text{keV}}

\newcommand{\Mb}{\text{Mb}}
\newcommand{\bq}{\begin{equation}}
\newcommand{\eq}{\end{equation}}
\newcommand{\bqn}{\begin{eqnarray}}
\newcommand{\eqn}{\end{eqnarray}}
\newcommand{\bfg}{\begin{figure}}
\newcommand{\efg}{\end{figure}}
 
\begin{document}

\title[Accurate photoionisation cross section for He]
      {Accurate photoionisation cross section for
       He at non-resonant photon energies}

\author{Alexander Stark$^1$ and Alejandro Saenz$^{1,2}$}
 \ead{alejandro.saenz@physik.hu-berlin.de}
\address{%
$^1$ AG Moderne Optik, Institut f\"ur Physik, Humboldt-Universit\"at zu Berlin,
                  Newtonstr.\,15, D\,--\,12\,489 Berlin, Germany\\
$^2$ Kavli Institute for Theoretical Physics, University of California, 
     Santa Barbara,\\
     CA 93106-4030, USA}%
\date{\today}

\begin{abstract}
The total single-photon ionisation cross section was calculated for  
helium atoms in their ground state. Using a full 
configuration-interaction approach the photoionisation 
cross section was extracted from the complex-scaled resolvent. 
In the energy range from ionisation threshold to 59\,eV our 
results agree with an earlier $B$-spline based calculation in which the 
continuum is box discretised within a relative error of $0.01\%$ 
in the non-resonant part of the spectrum. Above the $\He^{++}$  
threshold our results agree on the other hand very well to a recent 
Floquet calculation. Thus our calculation confirms the previously 
reported deviations from the experimental reference data outside the 
claimed error estimate. In order to extend the calculated spectrum 
to very high energies, an analytical hydrogenic-type model tail is 
introduced that should become asymptotically exact for infinite photon 
energies. Its universality is investigated considering also H$^-$, 
Li$^+$, and HeH$^+$. With the aid of the tail corrections to the 
dipole approximation are estimated.    
\end{abstract}


\maketitle

\section{Introduction}

Since the beginning of quantum mechanics the photoionisation cross section 
(PCS) of the helium atom was investigated in a number of experiments 
and numerical calculations. 
Helium is one of the simplest quantum mechanical systems that is relatively 
easily experimentally accessible, but also amenable to very accurate 
calculations. At the same time, it is theoretically challenging, since 
even within non-relativistic quantum mechanics the helium atom cannot 
be solved analytically. Due to this special characteristics the helium 
PCS is very attractive for a direct comparison of theory and experiment. 

In 1994 high-precision measurements of the PCS were performed 
by Samson \etal \cite{sfa:sams94,sfa:sams02} using a double ion chamber and
a high-voltage spark discharge. Since then the therein reported values for 
the PCS of helium with an estimated accuracy of $1-1.5\%$ in the low-energy 
range and about 2\,\% for the high-energy part beyond 
the double ionisation threshold were used in many applications in, 
e.\,g., astrophysics, plasma physics, and chemistry. Recently, the 
PCS of helium became also relevant for the characterisation of novel 
light sources like high-harmonic radiation or free-electron lasers (FEL). 
The supposedly very accurately known PCS of helium provides a natural 
way for tests and calibrations (especially of the intensity) of these 
new-generation light sources \cite{sfa:well08,naga07}. For example, in the
SASE (Self-Amplified Stimulated Emission) experiment at
FLASH (Hamburg) the two-photon double photoionisation of helium is used to
determine the duration of ultrashort femtosecond pulses \cite{mitz09}. 
Thereby, the nonlinear autocorrelation of direct 
$\He\rightarrow\He^{2+} + 2e^{-}$  
and sequential $\He\rightarrow\He^{+}+e^{-}\rightarrow\He^{2+}+2e^{-}$ 
double ionisation process is measured. The first step of the sequential 
process corresponds, of course, to single-photon ionisation of He. 

However, despite the development of new theoretical approaches and the 
access to increasing computational power it has so far not been possible 
to reproduce the experimental reference data in \cite{sfa:sams94,sfa:sams02} 
to the therein claimed accuracy. Supposedly very accurate theoretical 
values for the PCS of helium from the ionisation threshold to photon 
energies of 71\,eV were reported by Venuti \etal in \cite{bsp:venu96}. 
This was a follow-up work to the one presented in \cite{sfa:decl94} 
which extended to 2\,keV but was less accurate. 
The calculations were performed within the configuration-interaction 
approach in which the orbitals were expressed in $B$ splines (for the 
radial part) multiplied by spherical harmonics (for the angular part). 
The finite range of the adopted $B$ splines leads to a box-type 
discretisation for the continuum wave functions.   
From convergence studies and the agreement between the results obtained 
using the length, velocity, or acceleration form of the dipole operator 
the authors of \cite{bsp:venu96} estimated the error to be smaller 
than $0.001\,\Mb$ which corresponds to a relative error of 
$0.014-0.063\,\%$. In comparison to the experimental data in 
\cite{sfa:sams94,sfa:sams02} a deviation of up to about 2.6\,\% was, 
however, found in the considered low-energy range which is almost 
twice the estimated experimental error. 

More recently, Ivanov and Kheifets implemented a Floquet approach and 
calculated the PCS of helium starting at a photon energy of 80\,eV 
\cite{sfa:ivan06}. Again, the authors claim to reach an accuracy 
of the order of the fraction of a percent, but find noticeable deviations 
from the experimental reference data that easily reach up to 6\,\%. For 
a single value of the photon energy, 40\,eV, a good agreement was on the 
other hand found with 
the theoretical results in \cite{bsp:venu96}. As a consequence, 
the non-relativistic dipole and infinite-mass approximations 
used in both calculations appear to be inadequate or either the calculations 
or the experimental data are less accurate than claimed by the 
respective authors. In order to shed more light on this question, 
we performed calculations with a different theoretical approach 
in which the orbitals used in the subsequent configuration-interaction 
method are constructed from Slater-type orbitals. Furthermore, 
the PCS is extracted from the complex-scaled resolvent and thus 
no box discretisation as, e.\,g., in \cite{bsp:venu96} is used. 
We extend our {\it ab initio} results to very high photon energies 
by introducing an analytical, hydrogen-like model tail. This is  
of interest, for example, in view of the present efforts of extending the 
FEL sources to the x-ray regime, as with the LCLS (linac coherent 
light source) at Stanford, the XFEL (x-ray free electron laser) 
in Hamburg, or the SPring8-XFEL project in Japan. 
The universality of the here introduced tail is investigated by 
considering also the other two-electron 
systems H$^-$, Li$^+$, and even the molecular ion HeH$^+$. The tail is 
finally used to obtain an analytical estimate of the first-order 
correction to the dipole approximation.

\section{Method and Computational Details}

\subsection{Ab initio calculation}

In view of the large mass difference of the He nucleus and the electrons 
we adopt the infinite-mass approximation for the $\He$ nucleus 
in our calculations. This should be justified, since for the PCS of 
the $\He^+$ ion, for which the analytical result is known, 
the size of the modification due to the finite mass of the nucleus is 
only of the order of $0.1\%$. 
It is expected that this effect has a similar size in the case of the 
neutral $\He$ atom. Furthermore, there should not be a strong 
energy-dependent contribution due to the finite mass. In fact, for the ion 
it is energy independent.  
Relativistic effects should also be negligible, because their scaling 
parameter, the nuclear charge, is evidently a small value in the case 
of helium atoms. 

To obtain the wavefunctions and the corresponding 
energy eigenvalues of He we used a direct expansion in 
Slater-type orbitals (STOs). 
In the calculation of the \Nst{S} helium ground state 
the same basis-set parameters were used as in our recent calculation 
of the final-state spectrum of helium atoms after $\beta$ decay of 
tritium anions \cite{csm:star10}. 
The previous results for a large number of energy eigenvalues 
exhibited a very good agreement with literature values. 
For the \Nst{P} states we modified the \Nst{S} basis set as to adapt it 
to the different symmetry. As a result, we used 360 STOs leading to 
3331 configuration state functions (CSFs) for all \Nst{P} states and 555 
STOs resulting in 3481 CSFs for the \Nst{S} ground state. More details 
about the structure of the basis set and the construction 
of the symmetry-adapted CSFs used in the subsequent configuration-interaction 
(CI) calculation can be found in \cite{csm:star10}.

The photoionisation cross section $\sigma$ is related to the optical 
oscillator strength density ${\dd f}/{\dd E}$ by the 
relation \cite{sfa:fano68}
\bq
    \sigma =\frac{ \pi e^2 \hbar }{2 \epsilon_0 m_e c}
          \frac{\dd f(E)}{\dd E}
      =  109.7609\,\Mb\,\frac{\dd f(E)}{\dd E}\,\eV
\label{eq:Pint}
\eq
with the electron mass $m_e$ and charge $e$, 
the reduced Planck constant $\hbar$, the electric constant $\epsilon_0$,
and the speed of light in vacuum $c$.
Using atomic units ($m_e=1$, $e=1$, $\hbar=1$) the 
${\dd f}/{\dd E}$ can be evaluated from the dipole transition 
probabilities $P(E)$ 
\bq
  \left(\frac{\dd f(E)}{\dd E}\right)_{\au} = 2
  \cases{E P(E) & in length form \\ 
        {P(E)}/E & in velocity form.\\}
 \label{eq:oosdl}
\eq

Finally the $P(E)$ can be extracted 
from the complex-scaled resolvent according to \cite{csm:resc75,csm:saen03}
\bq
  P(E) = \frac{1}{\pi}\text{Im}\left\{
      \sum_{k}\frac{
              \bok{\Psi_i^{\rm S}(\theta^{*})}{\Op{d}(\theta)}
                  {\Psi_{k}^{\rm P}(\theta)}
              \bok{\Psi_{k}^{\rm P}(\theta^{*})}{\Op{d}(\theta)}
                  {\Psi_i^{\rm S}(\theta)}
                   }{E_k^{\rm P}(\theta)-E_i^{\rm S}(\theta)-E}
            \right\}\,.
\label{eq:P}
\eq
The complex-scaling angle is denoted by $\theta$ and 
$\bra{\Psi(\theta^{*})}$ is the biorthonormal eigenstate to 
$\ket{\Psi(\theta)}$. It is obtained from the latter by a transposition 
and complex conjugation of the angular part, while the radial part is only 
transposed but not conjugated. 
A variation of the angle $\theta$ allows the determination of an optimal 
angle $\theta_{\text{opt}}$ by requiring  
\bq
  \left.\frac{\partial P(E)}
             {\partial \theta}\right|_{\theta_{\text{opt}}}
         =\text{min.}
	\label{eq:opTh}
\eq
In other words, the best approximation of $P(E)$ is obtained with that value 
of $\theta$ for which $P(E)$ shows the smallest dependence on $\theta$. 
For more details about complex-scaling see 
\cite{csm:star10,csm:resc75,csm:saen93a} and 
references therein. An application of this method for the 
photoionisation of the molecular ion $\text{HeH}^{+}$ was described 
in \cite{csm:saen03}.

In \Eref{eq:P} the $\ket{\Psi_k^{\rm P}}$ are the final 
helium \Nst{P} wavefunctions and $E_k^{\rm P}$ the associated energy 
eigenvalues. Analogous definitions apply to the \Nst{S} initial state.  
The operator $\Op{d}$ describes the coupling of an atomic electron  
to a (classical) electromagnetic field. Thereby the latter is represented by 
a plane wave in the spatial domain, $e^{\mi \vek{k} \vek{r}}$, with 
the wave vector $\vek{k}$ and the spatial vector $\vek{r}$.
Without loss of generality we can assume the wave is propagating along 
the $z$ axis and the operator becomes in length and velocity forms  
\bq
  \Op{d}_l = - e^{\mi k \hat{z}}\hat{z}
   \label{eq:dl}
\eq
and
\bq
  \Op{d}_v = - e^{\mi k \hat{z}}\Op{v}_z \quad ,
 \label{eq:dv}
\eq
respectively. 
For sufficiently low energies the dipole approximation 
$e^{\mi k \hat{z}} \approx 1$ can be applied to a high level of 
reliability. The complex-scaled versions of the operators are simply 
given by  $\Op{d}_l(\theta) = - e^{+\mi\theta}\Op{z}$ and 
$\Op{d}_v(\theta) =  - e^{-\mi\theta}\Op{v}_z$. For an $N$-electron 
system the electronic part of the operator is simply the sum of 
the $N$ one-electron operators.

\subsection{Analytical model for high photon energies}

For very high photon energies we introduce an analytic tail. The concept 
behind its construction is that in the case of single-photon ionisation 
the ejected electron takes away a large fraction of the energy of the 
absorbed photon. Thus it escapes very fast from the nucleus and the 
remaining (spectator) electron. Therefore, the fast electron experiences 
the remaining system (with a maximum screening of the nucleus by the 
spectator electron) to a good approximation as a point 
charge $Z_f$ which is the sum of the charges of all remaining particles. 
(In the case of a $\He$ atom, the escaping electron experiences the 
remaining $\He^{+}$ ion as a point charge with $Z_f=+1$.) As a result, we may 
approximate the \Nst{P} two-electron wave 
function as (the spin part is omitted for better readability)
\bq
  \ket{\tilde{\Psi}^{1s\mathcal{E}p}(\mathcal{E})} = 2^{-\frac{1}{2}}
          \left[\ket{1s^{Z_i}\mathcal{E}p^{Z_f}}
               +\ket{\mathcal{E}p^{Z_f}1s^{Z_i}}\right] \quad .
  \label{eq:Psi_f}
\eq
Here, $\ket{\mathcal{E}p^{Z_f}}$ is the Coulomb continuum p-wave function
for energy $\mathcal{E}$ and charge $Z_f$, while $\ket{1s^{Z_i}}$ is a 
hydrogen-like ground-state wave function with effective charge $Z_i$. 
The initial \Nst{S} state is approximated as a product of two hydrogenic 
s orbitals with the same effective charge $Z_i$,
\bq
  \ket{\tilde{\Psi}^{1s^2}} = \ket{1s^{Z_i}1s^{Z_i}}.
  \label{eq:Psi_i}
\eq

In this model it is assumed that during the (fast) photoionisation 
process (escape of the emitted electron) the spectator electron has no 
time to relax. As a consequence, the effective charge $Z_i$ of the 
spectator electron remains the same in the initial \eref{eq:Psi_i} and 
the final \eref{eq:Psi_f} wave functions, but  
changes from $Z_i$ to $Z_f$ for the fast escaping electron.

Using this model we derive an analytical formula for the 
photoionisation cross section for two electron systems at high photon 
energies which reads in atomic units 
\bq
             \tilde{\sigma}_0^{L}(E) = N
\frac{512
\exp\left(-\frac{4 Z_f \arctan\left({\kappa}/{Z_i}\right)}{\kappa}\right)
       E Z_i^3 Z_f \left(Z_f-2 Z_i\right)^2  
      \left(Z_f^2+\kappa^2\right)}
     {3\left(1-\exp\left(-\frac{2 \pi  Z_f}{\kappa}\right)\right)
       \left(Z_i^2+\kappa^2\right)^6}
     \label{eq:tail}
\eq   
with $\kappa=\sqrt{2(E-I)}$, the photon energy $E$, the 
ionisation potential $I$, and the number of electrons $N$.
This expression is obtained when using the length-form representation 
of the dipole operator. For the velocity form of the dipole operator 
the model cross-section is
\bq
             \tilde{\sigma}_0^{V}(E) = 
                       \left(
                         \frac{Z_v\left(Z_v^2+\kappa^2\right)}
                              {2E \left(Z_f - 2 Z_v\right)}
                       \right)^2
                              \tilde{\sigma}_0^{L}(E)
     \label{eq:tailv}
\eq  
For the one-electron case ($N=1$,\,$Z_i=Z_v=Z_f$) the two representations 
become equal. This is, however, not the case for our two-electron model, 
since it adopts approximate wave functions. As a consequence, the 
corresponding commutator relation leading to equivalence between length 
and velocity forms of the dipole operator is not fulfilled. 

Within the model, two of the three parameters entering \Eref{eq:tail} 
and \Eref{eq:tailv}, $Z_f$ and $I$, are fixed by the sum of the 
charges of the remaining particles and the either experimentally 
or theoretically known ionisation potential, respectively. The choice 
of $Z_i$ ($Z_v$) is on the other hand less evident. One possibility is to 
relate it to the ionisation potential {\it via} $Z_i=\sqrt{2I}$ which 
leads to a parameter-free model tail \cite{sct:barb91,nu:saen97b}. 
A second choice is based on the 
requirement to yield the best possible ground-state description when 
adopting the trial wavefunction of \eref{eq:Psi_i} in a variational 
calculation. For He this results in the mean-field 
value $Z_i=1.6875$ \cite{beth77} compared to about $1.34$ for 
$Z_i=\sqrt{2I}$. A third alternative for the choice of $Z_i$ is 
to fit this parameter in order to provide the best agreement of the 
resulting tail with some {\it ab initio} or experimental PCS, within 
some energy interval. It should be noted that for $N=1$ and 
$Z_i=Z_f=\sqrt{2I}$ \Eref{eq:tail} is also evidently  
applicable for H-like systems and in this case one obtains the 
result given in \cite{beth77} and proposed as an approximate 
tail also for the generalised oscillator strength density 
in \cite{sct:barb91}. Clearly, as is discussed below for the example 
of HeH$^+$, also the present tail can be adopted to molecular 
systems. In the spirit of the Born-Oppenheimer approximation it 
may be useful to use in such a case an internuclear-separation 
dependent electron binding energy $I(R)$ instead of the experimental 
ionisation potential \cite{sfm:saen00c,sfm:saen00a,sct:luhr08b}. 

Through its simplicity this model offers the possibility to study 
analytically effects beyond the dipole approximation. 
Since the model becomes better for higher energy it is valid in the regime 
where the wavelength of the electric field may approach or even extends 
below the magnitude of 
$1\,\text{nm}$ and is thus comparable to inner-atomic distances. 
To estimate these effects in helium we derived expressions 
for the first-order corrections to the dipole approximation. 
In this case the one-electron interaction operator in length form is given by
\bq
  \Op{d}_l = -\Op{z}\exp(\mi k \Op{z}) 
          \approx  -\Op{z} - 
                   \mi E \alpha \Op{z}^2 
  \label{eq:d2}
\eq
with the photon energy $E=c \hbar k=({k}/{\alpha})$ (in a.\,u.) and the 
fine-structure constant $\alpha$. Since the initial state given in 
\Eref{eq:Psi_i} has $^1$S symmetry, the correction term 
$\mi E \alpha \Op{z}^2$ couples it only 
to final states with either $^1$S or $^1$D symmetry. 
In the spirit of the model one has then to substitute 
$\ket{\mathcal{E}p^{Z_f}}$ in \Eref{eq:Psi_f} with 
$\ket{\mathcal{E}s^{Z_f}}$ and $\ket{\mathcal{E}d^{Z_f}}$ for the 
continuum states with $^1$S and $^1$D symmetry, respectively. 

Since two distinct final states are reached at a given energy, the 
(incoherent) superposition of the cross-sections into the \Nst{S} and 
\Nst{D} channels is given by 
\bq
   \tilde{\sigma}^L_1(E) = \tilde{\sigma}^L_0(E)
        \left(1+\left(E \alpha\right)^2\left(A_s^2+A_d^2\right)\right)
\label{eq:tail1}
\eq
with
\bq
A_s =  \frac{3 Z_i \left(2 Z_f^2 -\kappa^2
               - Z_i\left(3 Z_f + Z_i\right)\right)
               + Z_f(2 \kappa^2 - Z_f^2)}
              {\sqrt{3} \left(Z_f - 2 Z_i\right) 
               \sqrt{Z_f^2+\kappa ^2} 
               \left(Z_i^2+\kappa ^2\right)}  
\eq
and
\bq
A_d = \frac{4 \left(Z_f-3 Z_i\right) 
              \sqrt{Z_f^2+4 \kappa ^2}}
             {\sqrt{15} \left(Z_f-2 Z_i\right) 
              \left(Z_i^2+\kappa^2\right)} \quad .
\eq
In the case of the \Nst{S} final states one obtains formally a term 
$\bk{\mathcal{E}s^{Z_f}}{1s^{Z_i}}\bok{1s^{Z_i}}{\Op{z}^2}{1s^{Z_i}}$
which can be interpreted as a shake up of the electrons.
It is an artifact from the unequal treatment of the two electrons in 
our model. Since in this case only one bound electron interacts 
with the electromagnetic field, the other electron (which will be ejected)
plays the role of the spectator and consequently experiences no 
relaxation. Thus to stay consistent within our model 
we have to exchange the effective charges in this term to 
$\bk{\mathcal{E}s^{Z_i}}{1s^{Z_i}}\bok{1s^{Z_f}}{\Op{z}^2}{1s^{Z_i}}$
which due to orthogonality of the $\ket{\mathcal{E}s^{Z_i}}$ and
$\ket{1s^{Z_i}}$ orbitals vanishes.

We successfully tested the derived expressions for the cross-sections 
in the one-electron case $Z=Z_i=Z_f$ with the aid of the sum 
rule \cite{aies:wang99}
\bq
 S_0 = \sum_n{2\left(E_n-E_0\right)
              \left|\bok{0}{\Op{z}^k}{n}\right|^2}
      + \int_I^{\infty}\tilde{\sigma}^L(E)\,dE 
     = k^2\bok{0}{\Op{z}^{2k-2}}{0}  \quad .
 \label{eq:sumrule}
\eq
In \Eref{eq:sumrule} the sum runs over all bound states $\ket{n}$ and  
$\ket{0}$ denotes the initial state, e.\,g., the ground state. 
In the case $k=1$ (dipole approximation) one has 
$\tilde{\sigma}^L=\tilde{\sigma}^L_0$
and the sum-rule result should be equal to the number of electrons. 
For $k=2$ (corrections) one finds $S_0={4}/{Z^2}$ by using 
$\tilde{\sigma}^L=\tilde{\sigma}^L_0(A_s^2+A_d^2)$.

\section{Results}
\subsection{Low-Energy Photoionisation Cross-Section}
\begin{table}
\caption{\label{tab:PCS} Total photoionisation cross sections $\sigma$ (in Mb) 
for $\He$ as a function of the photon energy $E$ (in eV). The present results are 
compared with experimental values of Samson \etal \cite{sfa:sams94} and with 
the  B-splines calculations of Venuti \etal \cite{bsp:venu96}.}
\begin{indented}
\lineup
\item[]\begin{tabular}{@{}lllllll}
\br
&\centre{2}{This work} & \centre{2}{Venuti \etal \cite{bsp:venu96}} & Samson \etal\cite{sfa:sams94}  \\
\ns
&\crule{2}&\crule{2}&\\

$E$	&    $\sigma_l$	&    $\sigma_v$ &    $\sigma_l$	&    $\sigma_v$	&
$\sigma_{\text{exp}}$ \\
\mr
24.596	&	7.38544	&	7.38369	&	7.39714	&	7.39676	&	7.40 \\
25	&	7.20885	&	7.20965	&	7.22037	&	7.22010	&	7.21 \\
26	&	6.79541	&	6.79524	&	6.79959	&	6.79946	&	6.79 \\
27	&	6.40743	&	6.40705	&	6.41103	&	6.41087	&	6.40 \\
28	&	6.04341	&	6.04224	&	6.04442	&	6.04412	&	6.05 \\
29	&	5.70323	&	5.70231	&	5.70546	&	5.70499	&	5.70 \\
30	&	5.38596	&	5.38555	&	5.38866	&	5.38802	&	5.38 \\
31	&	5.09035	&	5.08960	&	5.09260	&	5.09185	&	5.10 \\
32	&	4.81460	&	4.81401	&	4.81668	&	4.81586	&	4.82 \\
33	&	4.55779	&	4.55718	&	4.55960	&	4.55878	&	4.57 \\
34	&	4.31850	&	4.31751	&	4.31996	&	4.31918	&	4.32 \\
35	&	4.09519	&	4.09560	&	4.09642	&	4.09571	&	4.09 \\
36	&	3.88678	&	3.88722	&	3.88771	&	3.88707	&	3.88 \\
37	&	3.69217	&	3.69142	&	3.69277	&	3.69220	&	3.68 \\
38	&	3.51038	&	3.50966	&	3.51075	&	3.51024	&	3.50 \\
39	&	3.34037	&	3.33981	&	3.34078	&	3.34032	&	3.32 \\
40	&	3.18127	&	3.18072	&	3.18173	&	3.18129	&	3.16 \\ 
41	&	3.03227	&	3.03168	&	3.03260	&	3.03217	&	3.01 \\
42	&	2.89263	&	2.89335	&	2.80295	&	2.89251	&	2.86 \\
43	&	2.76174	&	2.76247	&	2.76217	&	2.76171	&	2.72 \\
44	&	2.63892	&	2.63966	&	2.63931	&	2.63882	&	2.60 \\
45	&	2.52360	&	2.52437	&	2.52395	&	2.52344	&	2.48 \\
46	&	2.41526	&	2.41607	&	2.41572	&	2.41519	&	2.38 \\
47	&	2.31341	&	2.31362	&	2.31387	&	2.31332	&	2.28 \\
48	&	2.21781	&	2.21778	&	2.21820	&	2.21764	&	2.19 \\
49	&	2.12792	&	2.12790	&	2.12830	&	2.12773	&	2.10 \\
50	&	2.04344	&	2.04346	&	2.04373	&	2.04317	&	2.02 \\
51	&	1.96412	&	1.96416	&	1.96442	&	1.96388	&	1.94 \\
52	&	1.88976	&	1.88986	&	1.88999	&	1.88046	&	1.85 \\
53	&	1.82026	&	1.82046	&	1.82049	&	1.81999	&	1.77 \\
54	&	1.75562	&	1.75596	&	1.75584	&	1.75537	&	1.71 \\
55	&	1.69624	&	1.69692	&	1.69639	&	1.69594	&	1.67 \\
56	&	1.64283	&	1.64347	&	1.64289	&	1.64247	&	1.63 \\
57	&	1.59740	&	1.59797	&	1.59734	&	1.59695	&	1.61 \\
58	&	1.56612	&	1.56668	&	1.56606	&	1.56569	&	1.58 \\
59	&	1.57722	&	1.57741	&	1.57656	&	1.57618	&	1.56 \\
\br
\end{tabular}
\end{indented}
\end{table}

\bfg
\centering
  \includegraphics[width=\columnwidth]{pcs_low_inset.eps}
\caption{\label{fig:pcs_low}
     (Colour online) Total photoionisation cross section of helium   
     (\Nst{S} ground state) from the ionisation 
     threshold to 59\,eV (red \full: present work, 
     $\times$: theoretical values of 
     Venuti \etal \cite{bsp:venu96}, 
     \opensquare: experimental values of Samson 
     \etal \cite{sfa:sams94,sfa:sams02}).
     The inset shows the theoretical PCS of this work and 
     experimental values of Samson \etal
     on a logarithmic scale for higher photon energies.}
\efg

In Table \ref{tab:PCS} the calculated {\it ab initio} 
PCS values are listed for the length and velocity forms of the dipole operator. 
A remarkable agreement of the order of $10^{-3}\text{Mb}$ can be 
noticed between these two formulations that for exact wave functions 
yield identical results. The results are given for the optimal values 
$\theta_{\text{opt}}$ obtained according to \Eref{eq:opTh}. However, 
overall, only a very small variation of the PCS with $\theta$ is found 
which indicates that the used basis set is rather complete for the 
considered energy range. 

\bfg
\centering
\includegraphics[width=\columnwidth]{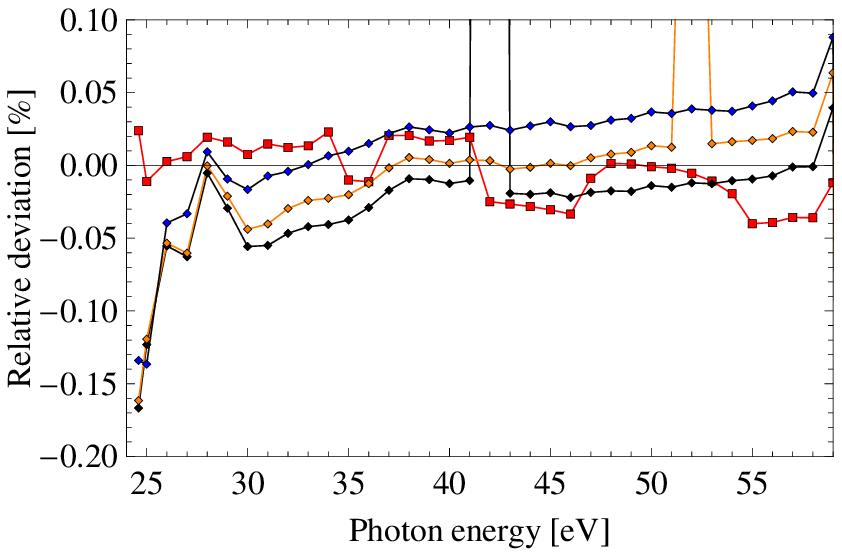}
\caption{\label{fig:err_own}(Colour online) 
         Relative deviation of our results in velocity form 
         (red \opensquare) and of the results of 
         Venuti \etal \cite{bsp:venu96} (\opendiamond) in 
         length (orange), velocity (black), and
         acceleration forms (blue) 
         from the helium photoionisation cross section 
         calculated in this work within the length formulation. 
         }
\efg

\bfg
\centering
\includegraphics[width=\columnwidth]{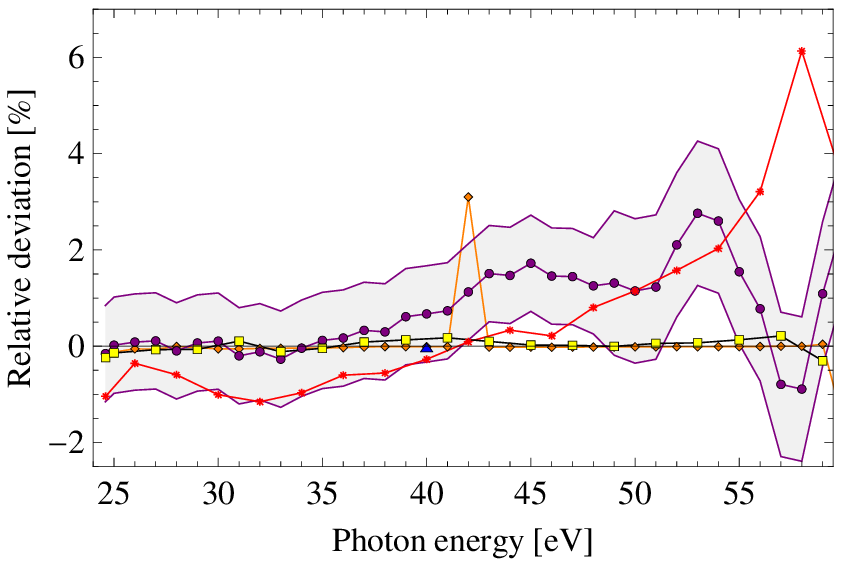}
\caption{\label{fig:err_low}(Colour online) Relative deviation from   
         the helium photoionisation cross section calculated in this 
         work (length form): theoretical data  
         of Venuti \etal \cite{bsp:venu96} (orange \opendiamond, length form),
         Chang and Fang \cite{bsp:chan95} (yellow \opensquare, length form),
         as well as Ivanov and Kheifets (single point at 40\,eV, 
         blue \opentriangle)
         \cite{sfa:ivan06}, the experimental values of
         Samson \etal \cite{sfa:sams94,sfa:sams02} (violet \opencircle), and
         the compiled data of Yan \etal \cite{sct:yan98} (red $\ast$). 
         The shaded area illustrates the error range estimated 
         by Samson \etal for their experiment ($\pm1\%$ until 48\,eV 
         and $\pm 1.5\%$ starting from 49\,eV).
         }
\efg

Table \ref{tab:PCS} also compares the present results with the experimental 
data of Samson \etal \cite{sfa:sams94} and the supposedly 
most accurate previous calculation of Venuti \etal 
\cite{bsp:venu96}. If shown graphically, as in Figure 
\ref{fig:pcs_low}, there are almost no visible differences between 
the various theoretical and experimental results. 
 
The very good agreement of the present results with the ones 
in \cite{bsp:venu96} is confirmed in Figure \ref{fig:err_own} 
that shows the relative deviation to the present length-form data 
for photon energies below 59\,eV. Despite the completely different 
theoretical approaches the agreement is remarkably good, since the 
deviation is less than $0.16\%$ and especially in the 
energy range from $38\,\eV$ to $58\,\eV$ the relative deviation is of the 
order of $\pm0.03\%$ (except for the values in \cite{bsp:venu96} at 
$42\,\eV$ (length form) and at $52\,\eV$ (velocity form), which we 
suggest to be typos). Figure \ref{fig:err_own} shows also the deviation 
between our results in length and velocity form. Again good agreement with 
deviations less than $\pm0.04\%$ is found.  In an earlier work, Chang and 
Fang adopted in \cite{bsp:chan95} a theoretical approach that is practically 
identical to the one used in \cite{bsp:venu96}, but a smaller basis set 
was used. 
Our results agree also well with the ones in \cite{bsp:chan95} (relative error 
of about $0.1\%$), but are
consistently in better agreement with the ones in \cite{bsp:venu96}. 
Clearly, on this level of accuracy convergence of the adopted basis set 
is finally decisive, while the two completely different approaches 
($B$ splines with box discretisation vs.\ Slater-type orbitals with 
complex scaling) appear to yield identical results within the achieved 
level of convergence. This indicates the correct and numerically stable 
implementation of both approaches. In fact, in the case of the single 
data point (at 40\,eV) given within the energy range shown in 
Figure \ref{fig:err_own}, the result of the Floquet calculation 
by Ivanov and Kheifets \cite{sfa:ivan06} agrees also within $0.03\%$ 
with our results and the one in \cite{bsp:venu96}. Therefore, three 
different theoretical approaches agree within an extremely small relative 
error. 
 
On the other hand, the experimental helium PCS of Samson \etal 
\cite{sfa:sams94,sfa:sams02} is still the primary reference for 
benchmarks and comparison, despite the fact that recent numerical 
calculations \cite{bsp:venu96,sfa:ivan06} yielded cross sections that 
differ from the experimental values outside the estimated experimental 
error bars. In view of the already discussed very good agreement with 
the theoretical data in \cite{bsp:venu96} (Figure \ref{fig:err_own}), 
we confirm the deviations to the experimental data in 
\cite{sfa:sams94,sfa:sams02} outside the error bars. This can clearly  
be seen in Figure \ref{fig:err_low}. For photon energies between 
$51\,\eV$ and $55\,\eV$ the relative deviation of the results of 
Samson \etal from our data and also from the ones of Venuti \etal is  
evidently above the error of $\pm1.5\%$ estimated by Samson \etal 
\cite{sfa:sams94,sfa:sams02}. The largest deviation of approximately $2.6\,\%$ 
in comparison to our values occurs at $53\,\eV$ and $54\,\eV$. These 
deviations are too large to be explained by a failure of the 
approximations adopted in the present calculation. Neither effects due 
to the finite size of the nucleus nor relativistic effects should be 
of a magnitude that is sufficient for explaining such a discrepancy 
that in addition would have to be strongly photon-energy dependent. 
As is discussed below on the basis of the derived analytical high-energy 
tail, also the consideration 
of non-dipole terms and thus a break-down of the dipole approximation 
yields corrections that are orders of magnitudes smaller than the 
ones required to find agreement between theory and experiment.  

In view of the in the Introduction discussed importance of the He PCS 
there exist also data sets that represent a compilation of experimental 
and theoretical data and try to cover large photon-energy ranges. 
Such a data set was reported by Yan \etal \cite{sct:yan98} and was 
claimed by its authors to be reliable for all energies. The comparison 
in Figure \ref{fig:err_low} shows that agreement with the here considered 
theoretical results is reasonable, but not really good. In fact, at both 
ends of the shown energy range the agreement of the theoretical data 
with the compiled ones is less good than the one found for the experimental 
data in \cite{sfa:sams94,sfa:sams02}. Since the compiled data lie below 
the theoretical ones for low energies and above for larger energies, 
the agreement is only good for intermediate energies close to the crossing 
point at about 42\,eV. Especially close to the ionisation threshold 
the experimental data in \cite{sfa:sams94,sfa:sams02} appear to be clearly 
superior to the compiled ones in \cite{sct:yan98}. In fact, within 
the first 10\,eV above the ionisation threshold the experimental data are 
in remarkable agreement to theory with a deviation of less than about 
$0.15\,\%$.

\subsection{High-Energy Photoionisation Cross-Section}
\begin{table}
\caption{\label{tab:PCSh} Total photoionisation cross sections $\sigma$ 
for $\He$ at high photon energies $E$ (in eV). Until $E=250\,$eV the $\sigma$ 
values are given in Mb and our results stem from the full {\it ab-initio} 
calculation. Starting from 400\,eV the results for the model 
tail \eref{eq:tail} are given and all $\sigma$ values are in Barn. 
Our results are compared to 
the experimental values of Samson \etal \cite{sfa:sams94}, 
the B-spline calculation of Decleva \etal \cite{sfa:decl94}, and the 
Floquet results of Ivanov and Kheifets \cite{sfa:ivan06}.}
\begin{indented}
\lineup
\item[]\begin{tabular}{@{}lllllll}
\br
&\centre{2}{This work} & \centre{2}{Decleva \etal \cite{sfa:decl94}} & Samson \etal\cite{sfa:sams94}  & Ivanov \etal \cite{sfa:ivan06}\\
\ns
&\crule{2}&\crule{2}&\\
$E$	&    \0\0\0$\sigma_l$	&    \0\0\0$\sigma_v$ &    \0\0\0$\sigma_l$	&    \0\0\0$\sigma_v$	&
\0\0\0$\sigma_{\text{exp}}$ & \0\0\0$\sigma_{\text{Fl}}$ \\\mr
\0\080      &    \0\0\00.73721 & \0\0\00.73777 &	\0\0\00.759   & \0\0\00.74    &       \0\0\00.693  & \0\0\00.7369 \\
\0\085      &    \0\0\00.63041 & \0\0\00.63099 &	\m---         & \m--- 	      &       \0\0\00.595  & \0\0\00.6308 \\
\0\091      &    \0\0\00.52708 & \0\0\00.52762 &	\m---         & \m--- 	      &       \0\0\00.502  & \0\0\00.5272 \\
\0\095      &    \0\0\00.47011 & \0\0\00.47063 &	\m---         & \m--- 	      &       \0\0\00.45   & \0\0\00.4701 \\
\0100     &    \0\0\00.40960 & \0\0\00.41004 &    \0\0\00.417   & \0\0\00.403   &       \0\0\00.393  &\m---\\
\0111     &    \0\0\00.30808 & \0\0\00.30854 &	\m---         &	\m--- 	      &       \0\0\00.3    & \0\0\00.3082 \\
\0120     &    \0\0\00.24809 & \0\0\00.24863 &    \0\0\00.251   & \0\0\00.243   &       \0\0\00.244  &\m---\\
\0140     &    \0\0\00.16041 & \0\0\00.16108 &    \0\0\00.167   & \0\0\00.158   &       \0\0\00.160  &\m---\\
\0160     &    \0\0\00.10924 & \0\0\00.10985 &    \0\0\00.112   & \0\0\00.108   &       \0\0\00.108  &\m---\\
\0180     &    \0\0\00.07749 & \0\0\00.07799 &    \0\0\00.802   & \0\0\00.077   &       \0\0\00.076  &\m---\\
\0205     &    \0\0\00.05270 & \0\0\00.05321 &	\m--- 	      &	\m--- 	      &       \0\0\00.051  & \0\0\00.0529 \\
\0250     &    \0\0\00.02881 & \0\0\00.02939 &    \0\0\00.0306  & \0\0\00.0293
&       \0\0\00.0277 &\m---\\
          & & & & & & \\
\0400     &       6561       &    6812       &    7270       &    7009       &    6370    &\m---\\
\0600     &    1785          &    1871       &    2001          &   1940        &    1770            &\m---\\
1000    &    \0335         &   \0354       &    \0395         &   \0384       &    \0339           &\m---\\
2000    &    \0\033.1      &   \0\035.2     &    \0\039.5      &   \0\040      &    \0\034.8        &\m---\\
3000    &       \0\0\08.42    &       \0\0\08.96    &      \m---      &     \m---       &      \0\0\08.77   &\m---\\ 
4000    &       \0\0\03.17    &       \0\0\03.38    &      \m---      &     \m---       &      \0\0\03.20   &\m---\\
5000    &       \0\0\01.48    &       \0\0\01.58    &      \m---      &     \m---       &      \0\0\01.47   &\m---\\
6000    &       \0\0\00.79    &       \0\0\00.85    &      \m---      &     \m---       &      \0\0\00.77   &\m---\\
7000    &       \0\0\00.47    &       \0\0\00.50    &      \m---      &     \m---       &      \0\0\00.45   &\m---\\
8000    &       \0\0\00.30    &       \0\0\00.32    &      \m---      &     \m---       &      \0\0\00.28   &\m---\\
\br
\end{tabular}
\end{indented}
\end{table}

Motivated by the recent work of Ivanov and Kheifets \cite{sfa:ivan06} 
who claim an even larger deviation from the experimental data of 
Samson \etal for larger photon energies than for the lower ones  
considered in \cite{bsp:venu96}, we also studied the single-photon 
ionisation process in the non-resonant energy regime above the $\He^{+}$ 
ionisation threshold ($E \approx 79\,\eV$). 

Although our approach, the expansion of the two-electron wave function in 
Slater-type orbitals, was originally developed for bound-state transitions, 
the extension by the complex-scaling method provides also an excellent 
description of high-energy continuum states of the $\He$ 
atom \cite{csm:star10}. 
Figure \ref{fig:err_high} shows the deviation of various theoretical, 
experimental, and compiled data from our results for energies between 
$80\,\eV$ and $205\,\eV$. The agreement with an older B-spline calculation 
by Decleva \etal \cite{sfa:decl94} is by far not as good as the 
one found with the later work \cite{bsp:venu96} of the same authors that 
concentrated on the low-energy regime. 
However, the deviation of our values from the Floquet 
results of Ivanov and Kheifets \cite{sfa:ivan06} is for most of the 
data points less or equal to about $0.03\%$. Only at the highest 
energy, $205\,\eV$, a deviation of $0.37\%$ is found. Therefore, we 
again confirm the discrepancy between previous theoretical calculations 
and the experimental reference data of Samson {\it et al}, this time  
reaching to about $6\%$ at $80\,\eV$, as can also be seen from the 
comparison of the different results in Table 2. 

On the other hand, we find in the energy range between about 110\,eV 
and 160\,eV a deviation from experiment that remains basically 
below $2\,\%$. Furthermore, the compiled data of Yan \etal \cite{sct:yan98} 
agree in this complete energy interval better with the theoretical results 
than the experimental results of Samson {\it et al}, in contrast to the 
findings at lower photon energies. In the energy interval between 
about 80\,eV and 100\,eV the experimental values reported by 
Bizau and Wuilleumier \cite{sfa:biza95} deviate from our calculation 
in a qualitatively very similar fashion as the experimental data 
of Samson {\it et al}. However, quantitatively, the deviation is smaller 
and remains in between 3 and $5\,\%$. In between 100\,eV and 150\,eV 
the data in \cite{sfa:biza95} are on the other hand substantially smaller 
than our theoretical results, leading to a deviation of up to $7\,\%$. 
For the higher energies shown in Figure \ref{fig:err_high} the 
deviation of both experimental data sets from our results is again 
qualitatively similar, but the ones of Bizau and Wuilleumier lie 
below our results and approach the latter with increasing energy, 
while the ones of Samson \etal lie above and thus agreement with 
our data becomes worse for increasing photon energy.   
 
\bfg
\centering
\includegraphics[width=\columnwidth]{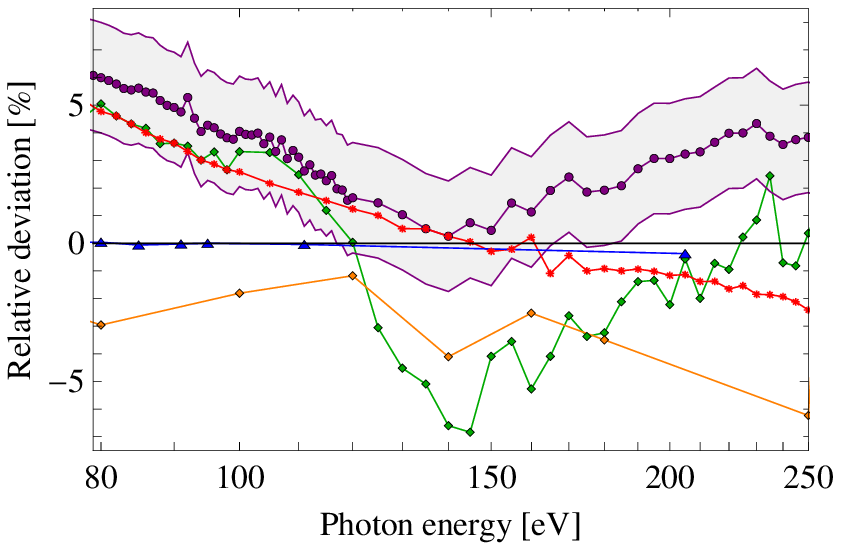}
\caption{\label{fig:err_high}(Colour online) Relative deviation from  
         the helium photoionisation cross section calculated in this 
         work: Floquet calculation of Ivanov and Kheifets \cite{sfa:ivan06} 
         (blue \opentriangle), 
         B-spline calculation of Decleva \etal \cite{sfa:decl94}
         (orange \opendiamond)
         experimental data measured by  
         Samson \etal \cite{sfa:sams94} (violet \opencircle) or 
         Bizau and Wuilleumier \cite{sfa:biza95} (green \opendiamond), 
         and the compiled values by Yan \etal \cite{sct:yan98} (red $\ast$). 
         The shaded area illustrates the error range estimated 
         by Samson \etal for their experiment ($\pm2\%$).
         }
\efg

For very high photon energies above $300\,\eV$ we propose to use the 
analytical model tail introduced in \Eref{eq:tail}.
In order to determine the tail parameter $Z_i$, we performed a 
least-squares fit of the tail to our \textit{ab initio} calculation in the  
energy range from $136\,\eV$ to $272\,\eV$ and obtained $Z_i=1.5293$. 
This value lies in between the prediction based on either 
$Z_i=\sqrt{2I}=1.34$ or the mean-field result, $Z_i=1.6875$.  
Although the fit interval spans only a small energy region of the 
calculated PCS, the extrapolated PCS agrees in a much larger 
energy range well with the \textit{ab-initio} results and in the complete 
shown energy range from 300\,eV to $8\,\keV$ very well with the 
experimental data of Samson \etal (Figure \ref{fig:velen}). 
In fact, we found almost the same effective initial charge, 
$Z_i=1.53\pm0.01$, when fitting the tail to the experimental data of 
Samson {\it et al}. The indicated uncertainty arises from different starting 
points of the energy span at the fitting procedure, while the end point 
was always chosen at an energy of $8\,\keV$.
Figure \ref{fig:pcs_high} shows two alternative tails besides the 
already discussed one. 
If $Z_i$ is fixed to its mean-field value (1.6875) and 
$Z_f$ is used as a fit parameter, the resulting tail lies above the 
experimental data of Samson {\it et al}. While the slope differs for 
photon energies at about 300\,eV, an almost constant off-set is 
found between the tail and the experimental data for large energies, 
if plotted on a doubly logarithmic scale. A further off-set is 
observed, if all tail parameters are fixed on the basis of simple 
arguments, i.\,e., the initial charge $Z_i$ is set to the mean-field 
value $1.6875$ and the final charge $Z_f$ to the value $1.0$ that 
should be appropriate at very large separation of the ejected electron. 
Not shown is the tail that is obtained under the assumption of no 
relaxation ($Z_i=Z_f$), since this tail does not agree to the 
experimental data at all. 

\bfg
\centering
\includegraphics[width=\columnwidth]{pcs_velen.eps}
\caption{\label{fig:velen} (Colour online)
     Analytical model tails for different representations of the dipole operator:
     length-form (red line) and velocity-form (blue line) for their optimal 
     parameters $Z_i$ and $Z_v$. Also the present {\it ab initio} results 
     and the 
     experimental data of Samson \etal are plotted. The inset shows the 
     curves on 
     a linear scale for low energies.}
\efg

\bfg
\centering
\includegraphics[width=\columnwidth]{pcs_high.eps}
\caption{\label{fig:pcs_high}(Colour online) The analytic tail   
         (see \Eref{eq:tail}, solid lines) that is an   
         approximation to the photoionisation 
         cross section of He at high energies is shown  
         for different parameters $Z_i$ and $Z_f$ 
         (as specified in the graph) and is 
         compared to the experimental values of Samson \etal 
         \cite{sfa:sams94} (violet \opencircle) and the calculation of 
         Decleva \etal \cite{sfa:decl94} (orange \opendiamond). 
         The dotted and dashed  
         lines show the first-order corrections to the dipole 
         approximation according to \Eref{eq:tail1} due to 
         final states with S or D symmetry, respectively. 
         The inset shows the high-energy part of the spectrum on a linear 
         scale.
         }
\efg

Alternatively to the described tail, an asymptotic representation of 
the cross section is often used as an estimate in the high-energy regime. 
The asymptotic limit of our length-form model tail is (in atomic units)
\bq
      \tilde{\sigma}_0^{L}(E) \approx N
            \frac{8 \sqrt{2} Z_i^3 (Z_f-2 Z_i)^2}
                 {3 \pi} E^{-7/2}
     \label{eq:atail}
\eq  
and for the photoionisation of helium we obtain
$\tilde{\sigma}_0^{\He}(E) \approx 487.956\,E\,(\keV)^{-7/2}$. The   
pre-factor is consistent with the values discussed by Samson \etal for 
different 
experimental data. It should be noted, however, that the analytic tail 
in \eref{eq:tail} possesses a much larger validity regime than the 
asymptotic expression in \eref{eq:atail}, 
since it shows good agreement with experiment and {\it ab initio} 
theory already for much lower photon energies. 

As mentioned above the length and velocity representations of the dipole 
operator lead to different expressions for the model tail. However, one 
may use the different representations to perform a consistency 
check of our model and to learn about the reliability of the fit 
procedure. For energies near infinity where the model is surely 
applicable the representations should lead to equivalent values. 
Thus we demanded the identity of the length and velocity model tails
in the first order of the asymptotic expansion. The asymptotic limit 
of the model tail in velocity representation is
\bq
      \tilde{\sigma}_0^{V}(E) \approx N
            \frac{8 \sqrt{2} Z_v^5}
                 {3 \pi} E^{-7/2} \quad .
     \label{eq:avtail}
\eq
Combining \eref{eq:atail} and \eref{eq:avtail} allows to define a 
fixed relation between $Z_i$ and $Z_v$ and thus to define a new 
parameter $Z_v^*$ that fulfils this relation, 
\bq
     Z_v^*=\left(4 Z_i^5-4 Z_f Z_i^4+Z_f^2 Z_i^3\right)^{1/5} \quad .
     \label{eq:Zv}
\eq
For $Z_i=1.5293$ and $Z_f=1$ this expression leads to $Z_v^*=1.7224$, while by 
applying the same fit procedure for $\tilde{\sigma}_0^{V}$ as was used 
before for 
$\tilde{\sigma}_0^{L}$ we find $Z_v=1.7455$. Clearly, the differently 
obtained values for $Z_v$ and $Z_v^*$ are in quite reasonable agreement. 
Thus we conclude that the model tail, although it is derived 
with approximate wave functions, yields for very high energies the 
required independence of the chosen representation (length or 
velocity form) of the dipole operator. Furthermore,  
because the results of \Eref{eq:Zv} and the fit are in very good 
agreement, it is evident that the fit procedure is most suitable to derive 
the parameter values for the model tail. 
It may finally be observed that $Z_v$ (and $Z_v^*$) are close to the mean-field 
value 1.6875. This may indicate that, if one would like to avoid the fit 
or if no data are available for performing a fit, the best choice for 
a parameter-free tail is the tail in velocity form adopting the 
mean-field prediction for $Z_v$. 

In Figure \ref{fig:velen} the  
length and velocity model tails are plotted for their optimal 
values. It can be observed that the photon energy where the model tail 
approaches the {\it ab initio} calculation is lower for the velocity form 
than for the length form. In fact, it is quite surprising how well the 
velocity-form tail agrees to the full {\it ab-initio} results and the 
experimental data even down to the ionisation threshold. This 
superiority of the velocity form compared to the length form found 
at low energies is, however, on the first glance a little bit surprising, 
since the transition dipole matrix elements in length form are usually 
supposed to be preferable at low photon energies \cite{sfa:ivan06,john64}. 
This is often explained by the lower sensitivity of the length-form 
matrix elements to errors in the long-range part of the wavefunctions 
and the fact that low-energy transitions are more sensitive to this  
long-range part. However, it may be remembered that the obtained value 
of $Z_v$ is quite close to the mean-field prediction. Since the low-energy 
part of the photoionisation spectrum is more sensitive to the details of 
the atomic potential, a model like the velocity-form tail that is closer 
to the mean-field prediction may thus be favourable compared to the 
length-form tail with a rather different value found for the 
corresponding parameter $Z_i$.  

In order to demonstrate the generality of the analytical tail, 
we investigated its applicability to other two-electron systems. 
Besides the neutral helium atom we considered the hydrogen anion, 
the lithium cation, and even the {\HeH}\ molecular cation. The used 
tail parameters are listed in Table \ref{tab:Tail}. 
Figure \ref{fig:tails} shows a comparison of the tails for these 
systems with corresponding literature data and, for the atoms, also 
with {\it ab-initio} results calculated in this work. The found agreement is 
in all cases good or even very good. This indicates the universality 
of the tail concept. It is, in fact, interesting that the tail works well 
even for a system like $\text{H}^{-}$ in which the escaping electron 
experiences within the tail model no influence from the remaining hydrogen 
atom, since polarisation effects are ignored. Furthermore, the 
initial state of H$^-$ is only bound due to correlation and thus 
a mean-field model is in principle not applicable. Nevertheless, 
also in this case the tail seems to work well. In the case of HeH$^+$ 
a high-energy tail had been proposed before \cite{csm:saen03}, but 
a paramter-free tail (with $Z_i=Z_f=\sqrt{2I})$ had been used. 
The result is also shown in Figure \ref{fig:tails}. It also compares 
reasonably well with the full ab-initio results. The reason is that 
in this case $Z_i$ is very close to $Z_f$ and, in fact, also to 
$\sqrt{I}$. Note, for a molecular 
system there is the additional complication due to nuclear motion. 
However, in the spirit of the high-energy tail it should have a 
negligible influence on the high-energy part of the photoionisation 
spectrum. Therefore, the tail is formally obtained for a single 
internuclear separation, usually the equilibrium distance. Here, 
we chose in agreement to \cite{csm:saen03} the ionisation energy 
$I=45\,$eV.    

\begin{table}
\caption{\label{tab:Tail} Parameters for the model tail for 
         various one- and two-electron systems.}
\begin{indented}
\item[]\begin{tabular}{@{}l|lllllll}
\br
Parameters   & $\text{H}^{-}$ & H        & $\text{He}$ & $\text{He}^{+}$ & $\text{Li}^{+}$ & $\text{HeH}^{+}$ \\\br
$Z_i$        & 0.5586         & 1        & 1.5293      & 2               & 2.5191          & 1.9414 \\
$Z_v$        & 0.7662         & 1        & 1.7455      & 2               & 2.7390          & 1.9026 \\
$Z_v^{*}$    & 0.7371         & 1        & 1.7224      & 2               & 2.7152         & 1.9177 \\
$Z_f$        & 0              & 1        & 1           & 2               & 2               & 2 \\
$N$          & 2              & 1        & 2           & 1               & 2               & 2 \\
$I$ [$\eV$]  & 0.75436        & 13.606   & 24.5912     & 54.4234         & 75.64           & 45 \\
\br
\end{tabular}
\end{indented}
\end{table}

\bfg
\centering
\includegraphics[width=\columnwidth]{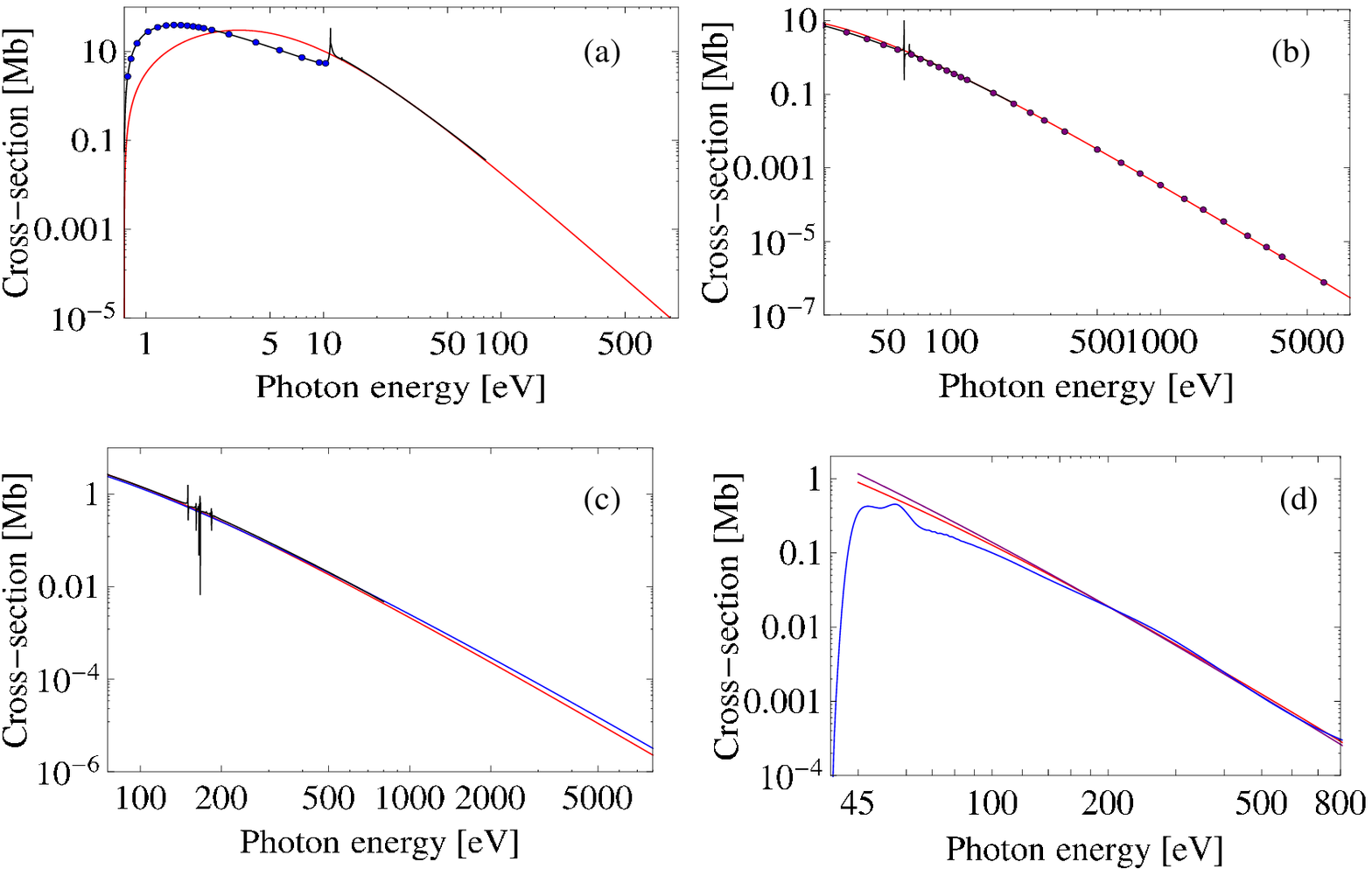}
\caption{\label{fig:tails} (Colour online)
     Analytical model tails (length form, red solid lines) for different 
     two-electron systems are compared to literature data.    
     a) $\text{H}^{-}$: Venuti and Decleva \cite{sfa:venu97} (blue \opencircle),
     b) $\He$: Samson \etal \cite{sfa:sams94} (violet \opencircle),
     c) $\text{Li}^{+}$: Verner \etal \cite{sfa:vern96} (blue line), and 
     d) $\text{HeH}^{+}$ (parallel contribution): 
        Saenz \cite{csm:saen03} (blue line). 
     For the atoms (a to c) also {\it ab-initio} cross sections 
     obtained within this work are shown (black solid lines), while 
     for HeH$^+$ the alternative tail proposed in \cite{csm:saen03} 
     is additionally given (purple solid line).}
\efg

According to \eref{eq:tail1} the model tail provides the 
possibility to give an analytical estimate of the first-order 
correction to the dipole approximation. The two contributions 
from final states with S and D symmetry are also shown in 
Figure \ref{fig:pcs_high}. While the D contribution is clearly 
dominant, the total correction to the dipole approximation is 
still negligibly small at photon energies of several keV. 
Note, however, that the relative contribution reaches already 
$11\%$ at $8\,\keV$ due to the very small cross section  
of less than $1\,\text{barn}$. This allows to conclude that 
very accurate studies of the photoionisation cross section 
of He performed in the keV photon energy range must take 
corrections to the dipole approximation into account. 
However, since the correction is proportional to the squared 
photon energy and the cross section itself increases by almost 
five orders of magnitude when going from 8\,keV to 300\,eV 
photons, corrections due to a break-down of the dipole 
approximation are negligible within the here considered level 
of accuracy for photon energies below 200\,eV. Therefore, the 
deviations between the theoretical results of the present 
work as well as the ones in \cite{sfa:venu97} or \cite{sfa:ivan06} 
from the experimental reference data of 
Samson \etal \cite{sfa:sams94,sfa:sams02} or Bizau and 
Wuilleumier \cite{sfa:biza95} cannot be explained by a failure 
of the dipole approximation.

\section{Summary}
An {\it ab initio} calculation of the photoionisation cross section 
of He has been performed for photon energies covering the non-resonant 
part of the spectrum from the ionisation threshold until about 
300\,eV. An analytical high-energy model tail has been introduced 
and the {\it ab initio} data were used in order to determine the 
single fit parameter. With this tail it became possible to predict 
the photoionisation spectrum for arbitrarily large photon energies 
within the underlying non-relativistic dipole approximation. 
Furthermore, the first-order correction to the dipole approximation 
could be estimated analytically with the aid of the model tail. 

Our theoretical results agree extremely well with the ones obtained 
by different theoretical approaches in the low and the high energy 
parts by Venuti \etal \cite{bsp:venu96} and Ivanov and 
Kheifets \cite{sfa:ivan06}, respectively. Therefore, we confirm 
the pronounced deviation between theoretical and experimental 
results noted in those earlier works. Particularly, at the 
photon-energy range around $50\,$eV that is very relevant to 
present-day FELs like FLASH or high-harmonic sources    
the relative deviation is unambiguously larger than the error 
estimates of the experiment of 
Samson \etal \cite{sfa:sams94,sfa:sams02}. Since we also 
demonstrated the validity of the dipole approximation at least for 
photon energies up to 200\,eV where the previous comparisons 
were performed, this possible source of disagreement between 
theory and experiment is excluded. We thus conclude that the 
quality of theoretical calculations of the helium photoionisation 
cross section has reached a consistently higher level than the 
experiment.  In view of the fundamental importance of the 
photoionisation cross section of helium we hope that the present 
work stimulates future experimental efforts to resolve the 
discrepancies between theory and experiment. Until this has 
been achieved, we propose to use the theoretical results, 
especially the ones of the present work that cover a large 
photon energy range in a consistent fashion, instead 
of the experimental ones as reference data for, e.\,g., 
calibrating new light sources, since they appear 
to be more accurate and reliable.

\ack
The authors acknowledge financial support from the {\it COST 
programme CM0702} and the {\it Fonds der Chemischen Industrie}. 
This work was supported in parts by the National Science Foundation 
under Grant No.\ NSF PHY05-51164. 

\bibliographystyle{iopart-num}

\section*{References}
 

\begin{thebibliography}{10}
\expandafter\ifx\csname url\endcsname\relax
  \def\url#1{{\tt #1}}\fi
\expandafter\ifx\csname urlprefix\endcsname\relax\def\urlprefix{URL }\fi
\providecommand{\eprint}[2][]{\url{#2}}
%

\bibitem{sfa:sams94}
Samson J~A~R, He Z~X, Yin L and Haddad G~N 1994 {\em J.\,Phys.\,B\/} {\bf 27}
  887

\bibitem{sfa:sams02}
Samson J~A~R and Stolte W~C 2002 {\em
  J.\,Electr.\,Spectros.\,Relat.\,Phenom.\/} {\bf 123} 265

\bibitem{sfa:well08}
Wellh\"{o}fer M, Hoeft J~T, Martins M, Wurth W, Braune M, Viefhaus J, Tiedtke K
  and Richter M 2008  {\bf 3} P02003

\bibitem{naga07}
Nagasono M, Suljoti E, Pietzsch A, Hennies F, Wellh\"ofer M, Hoeft J~T, Martins
  M, Wurth W, Treusch R, Feldhaus J, Schneider J~R and F\"ohlisch A 2007 {\em
  Phys.\,Rev.\,A\/} {\bf 75} 051406

\bibitem{mitz09}
Mitzner R, Sorokin A~A, Siemer B, Roling S, Rutkowski M, Zacharias H, Neeb M,
  Noll T, Siewert F, Eberhardt W, Richter M, Juranic P, Tiedtke K and Feldhaus
  J 2009 {\em Phys.\,Rev.\,A\/} {\bf 80} 025402

\bibitem{bsp:venu96}
Venuti M, Decleva P and Lisini A 1996 {\em J.\,Phys.\,B\/} {\bf 29} 5315

\bibitem{sfa:decl94}
Decleva P, Lisini A and Venuti M 1994 {\em J.\,Phys.\,B\/} {\bf 27} 4867

\bibitem{sfa:ivan06}
Ivanov I~A and Kheifets A~S 2006 {\em Eur.\,Phys.\,J.\,D\/} {\bf 38} 471

\bibitem{csm:star10}
Stark A and Saenz A 2010 {\em Phys.\,Rev.\,A\/} {\bf 81} 032501

\bibitem{sfa:fano68}
Fano U and Cooper J~W 1968 {\em Rev.\,Mod.\,Phys.\/} {\bf 40} 441

\bibitem{csm:resc75}
Rescigno T~N and McKoy V 1975 {\em Phys.\,Rev.\,A\/} {\bf 12} 522

\bibitem{csm:saen03}
Saenz A 2003 {\em Phys.\,Rev.\,A\/} {\bf 67} 033409

\bibitem{csm:saen93a}
Saenz A, Weyrich W and Froelich P 1993 {\em Int.\,J.\,Quant.\,Chem.\/} {\bf 46}
  365

\bibitem{sct:barb91}
Barbieri R~S and Bonham R~A 1991 {\em Phys.\,Rev.\,A\/} {\bf 44} 7361

\bibitem{nu:saen97b}
Saenz A and Froelich P 1997 {\em Phys.\,Rev.\,C\/} {\bf 56} 2162

\bibitem{beth77}
Bethe H~A and Salpeter E~E 1977 {\em Quantum Mechanics of One- and Two-Electron
  Atoms\/} (New York: Plenum)

\bibitem{sfm:saen00c}
Saenz A 2000 {\em J.\,Phys.\,B\/} {\bf 33} 4365

\bibitem{sfm:saen00a}
Saenz A 2000 {\em Phys.\,Rev.\,A\/} {\bf 61} 051402(R)

\bibitem{sct:luhr08b}
L\"uhr A, Vanne Y~V and Saenz A 2008 {\em Phys.\,Rev.\,A\/} {\bf 78} 042510

\bibitem{aies:wang99}
Wang S 1999 {\em Phys.\,Rev.\,A\/} {\bf 60} 262

\bibitem{bsp:chan95}
Chang T~N and Fang T~K 1995 {\em Phys.\,Rev.\,A\/} {\bf 52} 2638

\bibitem{sct:yan98}
Yan M, Sadeghpour H~R and Dalgarno A 1998 {\em Astrophys.\,J.\/} {\bf 496} 1044

\bibitem{sfa:biza95}
Bizau J~M and Wuilleumier F~J 1995 {\em
  J.\,Electr.\,Spectros.\,Relat.\,Phenom.\/} {\bf 71} 205

\bibitem{john64}
Johnston R~R 1964 {\em Phys. Rev.\/} {\bf 136} A958

\bibitem{sfa:venu97}
Venuti M and Decleva P 1997 {\em J.\,Phys.\,B\/} {\bf 30} 4839

\bibitem{sfa:vern96}
Verner D~A, Ferland G~J, Korista K~T and Yakovlev D~G 1996 {\em
  Astrophys.\,J.\/} {\bf 465} 487

\end{thebibliography}

\providecommand{\newblock}{}

\end{document}